# Modeling Agglomeration of Dust Particles in Plasma

Lorin S. Matthews, Victor Land, Qianyu Ma, Jonathan D. Perry, and Truell W. Hyde

*Center for Astrophysics, Space Physics, and Engineering Research*
*Baylor University, One Bear Place 97310, Waco, Texas 76798-7310 USA*

**Abstract.** The charge on an aggregate immersed in a plasma environment distributes itself over the aggregate's surface; this can be approximated theoretically by assuming a multipole distribution. The dipole-dipole (or higher order) charge interactions between fractal aggregates lead to rotations of the grains as they interact. Other properties of the dust grains also influence the agglomeration process, such as the monomer shape (spherical or ellipsoidal) or the presence of magnetic material. Finally, the plasma and grain properties also determine the morphology of the resultant aggregates. Porous and fluffy aggregates are more strongly coupled to the gas, leading to reduced collisional velocities, and greater collisional cross sections. These factors in turn can determine the growth rate of the aggregates and evolution of the dust cloud. This paper gives an overview of the numerical and experimental methods used to study dust agglomeration at CASPER and highlights some recent results.

**Keywords:** dusty plasma, coagulation, dust charging

**PACS:** 52.27.Lw, 94.05.Bf, 95.30.Wi, 96.30.Vb

## INTRODUCTION

Dusty plasmas can be found in nebular clouds, the clouds surrounding developing protostars and protoplanets, the ephemeral rings around planets, in cometary tails, and in planetary or lunar atmospheres [1-3]. Dusty plasmas are also a byproduct of the plasma processing of silicon wafers for computer chips and can be found in nuclear fusion reactors [4,5]. In each of these environments the dust particles can agglomerate, which changes their interaction with the plasma and affects their subsequent dynamics as well as the evolution of the system. The microphysics underlying these processes is complex, depending on parameters specific to the ambient environment and the grains themselves. Grains within a dusty plasma become charged due to the direct plasma current to the grain, with their equilibrium charge being affected by the distribution of the plasma velocities, usually assumed to be Maxwellian (in laboratory environments) or Lorentzian (as found in many astrophysical environments) [6,7]. Additional charging processes such as secondary electron emission or photoemission can lead to opposite charges on the monomers comprising an aggregate or to the aggregate charge becoming unstable and changing sign [8]. In addition to the plasma environment, special properties of the coagulating monomers, such as ferromagnetic material or non-spherical shapes, also play a role in the evolution of the agglomeration of aggregates. Recently, dust aggregates have been created in a laboratory plasma environment [9,10], which allows a test of the validity of numerical models and provides new insights to the interactions between charged dust grains.

## NUMERICAL MODELING OF AGGLOMERATION

Particles in a radiative or plasma environment will become charged due to the direct collection of plasma particles or through electron emission. The charge is distributed over the surface of the aggregate, which requires a numerical simulation to determine the charge on a highly irregular aggregate. A numerical model based on OML (orbital motion limited) theory modified to determine the open lines of sight (LOS) to points on the surface of the aggregate is used to calculate the charge and arrangement of charge on the aggregates. This model is used in conjunction with numerical codes which calculate the multipole interactions between colliding aggregates to self-consistently model the charging and growth of fluffy aggregates. The models also can allow for other special monomer properties, such as the magnetic interactions between ferromagnetic material or the effects of non-spherical grains.

### Charging of Aggregates

The charging of a single particle immersed in plasma is described by OML theory, originally derived for Langmuir probe measurements [6]. The current density due to incoming particle species α (here we assume α = e or α = + for electrons and ions carrying one positive electron charge, respectively) to a point on the surface of a particle is given by

$$J_\alpha = n_{\alpha\infty} q_\alpha \int_{v_{\min,\alpha}}^{\infty} f_\alpha v_\alpha^3 dv_\alpha \int \cos(\theta) d\Omega. \tag{1}$$

with $n_\alpha$ the plasma density very far from the particle, $q_\alpha$ the charge of the incoming plasma particle, $f_\alpha$ the velocity distribution function of the plasma particles, and $v_\alpha \cos(\theta)$ the velocity component of the incoming plasma particle perpendicular to the surface.

By specifying a plasma distribution function appropriate for the given environment, the integration over velocity can be readily carried out. The calculation of charge on a fractal aggregate is then reduced to determining the solid angle, $d\Omega$, for unobstructed orbits to each monomer within the aggregate. A LOS approximation is used to determine which orbits to exclude from the limits of integration [11]. The surface of the monomer is divided into equal-area patches. Vectors pointing from the center of the monomer to these surface points define test directions which are determined to be *blocked* if they intersect any other monomer in the aggregate, or the monomer in question, and *open* otherwise. Each line of sight is then assigned a value of 0, or 1, respectively. The number of 1's divided by the total number of test-directions comprises the *open lines of sight factor* for the patch. At the same time, the cosine of the angles between the normal direction of the patch and each test direction is also determined and then included in the integral. The change in charge of the monomer is then obtained by adding up the contribution of all the patches. The change in charge of the aggregate as a whole is obtained by adding the contribution from each of the N monomers. This process is iterated in time until the change in aggregate charge becomes negligible, dQagg < 0.0001%, at which point on average the net current to the aggregate will be zero.

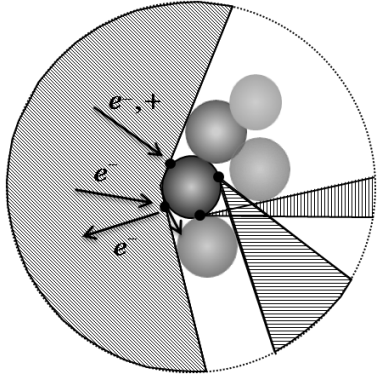

**FIGURE 1.** A 2D representation of the OML LOS geometry. Different points on a monomer in an aggregate are partly shadowed from the outside plasma or radiation by other monomers. The dashed areas indicate the open, unblocked lines of sight for four points on a monomer. In the case of electron emission, high energy electrons or photons can knock off an electron, which may escape the aggregate surface along an open line of sight, or be recaptured by another monomer in the aggregate, altering the arrangement of charge on the surface.

A similar treatment can be used to determine the flux of emitted electrons, arising from high-energy electrons or UV radiation. Figure 1 illustrates this method.

An interesting result is that even with only electron and ion plasma currents, a negatively charged aggregate can have positively charged monomers within it (Fig. 2a). This increases the dipole attractions between particles, enhancing coagulation. A second interesting result for aggregates is that charging by UV radiation can produce a "flip-flop" phenomenon similar to that discussed by Meyer-Vernet [8], but caused by photoemission rather than secondary electron emission (Fig. 2b). In contrast, single spheres have only one stable equilibrium potential when exposed to UV radiation.

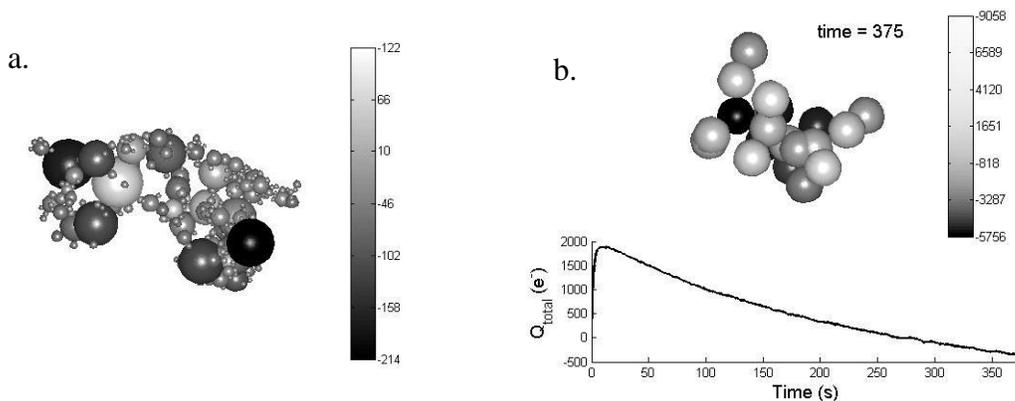

**FIGURE 2.** Charging of aggregates. a) Aggregate charged with only plasma currents. Though the overall charge on the aggregate is negative, some of the monomers within the aggregate are charged positively. b) Aggregate charged using the plasma parameters and UV flux at 1 AU from the sun. The charge is unstable in time and changes polarity.

## Collision Dynamics

Two numerical codes are used to study the collision dynamics of charged aggregates based on box_tree, an N-body code originally developed to investigate the gravitational interactions between planetesimals and objects in rocky rings [12]. The box_tree code was modified and extended to include the effects of charged particles and magnetic fields and treats accelerations caused by interactions of charged grains as well as rotations induced by torques due to the charge dipole moments [13,14]. These dipole-dipole interactions have been shown to greatly enhance the collision rate, even for like-charged particles [15].

Aggregate Builder is a subset of the box_tree code which is used to study pairwise interactions of colliding particles in the COM-frame of the target particle. The incoming particle has a velocity directed towards the target particle to within an offset-distance $(a_t + a_i)/2$ of the COM, where $a_t$ and $a_i$ are the maximum radii of the target and incoming particles, respectively. Libraries of aggregates are created from successful collisions, which can then be used as starting points in the box_tree code.

## Characterizing Aggregates

The fluffiness of an aggregate is very important to the coagulation process, since a fluffier aggregate couples more effectively to the gas in PPDs than does a compact aggregate [16]. The compactness factor [17] can be calculated by the ratio of the volume of the monomers within the aggregate to the volume of a sphere with a radius $R_\sigma$. $R_\sigma$ is the radius of a circle with area equal to the projected cross-section of the aggregate averaged over many orientations.

$$\phi_\sigma = \frac{\sum r_i^3}{R_\sigma^3} \qquad (2)$$

For fluffy aggregates, $\phi_\sigma$ approaches 0, for compact aggregates it approaches 1. An illustration for a representative aggregate is shown in Figure 3.

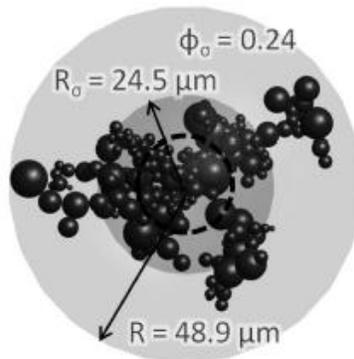

**FIGURE 3.** An illustration of the compactness factor for an aggregate. The outer circle indicates the maximum aggregate radius, R, defined as the maximum extent of the aggregate from the COM. The darker inner circle indicates $R_\sigma$.

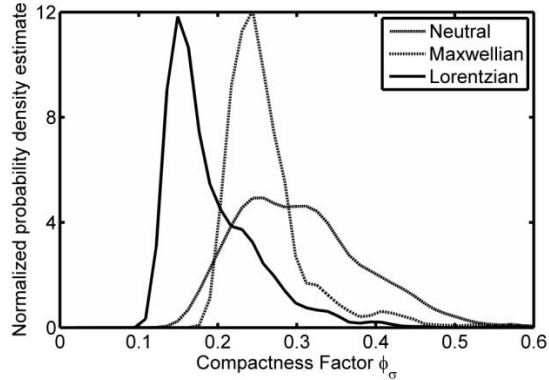

**FIGURE 4.** Probability density estimate for charged and uncharged aggregates. Charged aggregates tend to have lower compactness factors and the distribution tends to be sharply peaked.

The compactness factor of the aggregates is strongly influenced by the charge on the aggregates. Figure 4 compares the distribution of compactness factor for aggregates with 200 < N < 2000 monomers. The charged grains were exposed to plasma with the same temperatures and densities, but of the plasma distributions had a Maxwellian distribution for the electrons and the other had a Lorentzian distribution with κ = 5. The high energy tail of the Lorentzian distribution leads to more highly charged aggregates which in general had lower compactness factors, with the distribution among the population being sharply peaked. In contrast, neutral aggregates have a broad distribution of compactness factors within the population.

## Beyond Charged Spheres

Special characteristics of the monomers will also influence the morphology of the resultant aggregates. Monomers which consist of ferromagnetic material will experience magnetic dipole-dipole interactions. Interacting grains tend to align the dipoles so that open filamentary structures are formed, freezing in the magnetic moment [18]. Charged magnetic material will form aggregates with structural characteristics intermediate to purely charged or purely magnetic material, as shown in Figure 5. The shape of the monomers is also an important consideration. Ellipsoidal monomers tend to form aggregates which are less compact, but with much greater variation among the population (Figure 6).

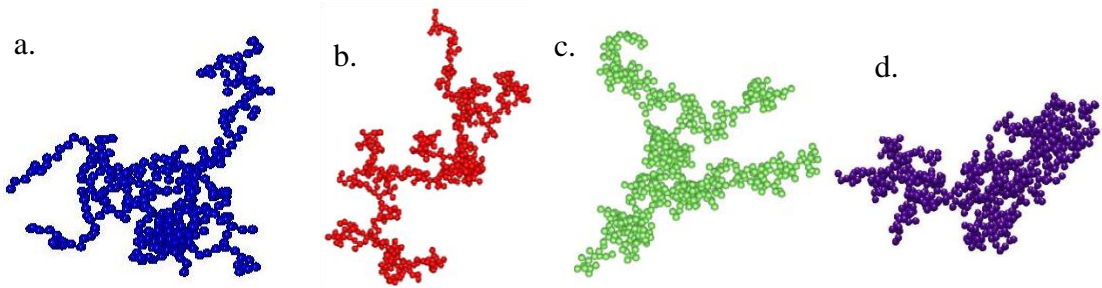

**FIGURE 5.** Aggregates formed from a) magnetic monomers b) charged-magnetic monomers c) charged monomers and d) neutral monomers.

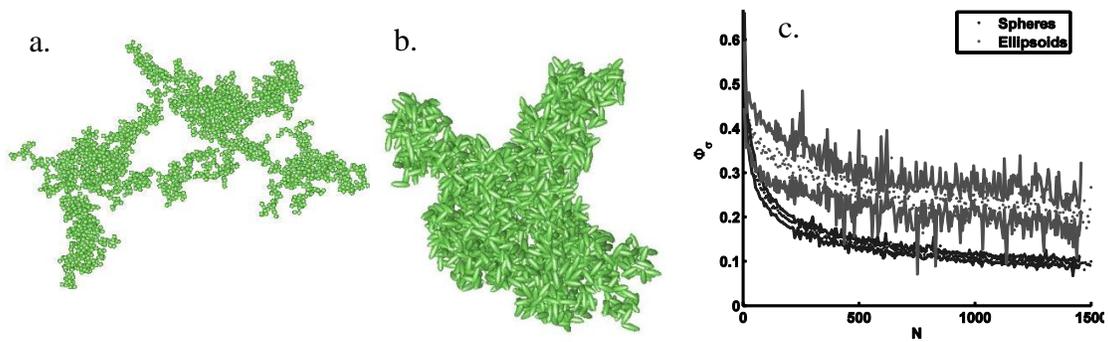

**FIGURE 6.** Aggregates made of a) spherical monomers and b) ellipsoidal monomers. Each aggregate contains about 2000 constituent monomers. c) The ellipsoidal monomers (gray points) form more compact aggregates than spheres (black points), but with a large variation for a given aggregate size.

## CONCLUSIONS

The non-uniform distribution of charge on dust aggregates leads to interesting dynamics within dust populations. Charging models show that monomers within the same aggregate can have opposite charges, and photoemission can lead to unstable charges on aggregates with a "flip-flop" in potential. Dynamic simulations show that the dipole-dipole interactions lead to enhanced coagulation, with charged aggregates being fluffier than neutral aggregates. Populations of charged aggregates tend to also build self-similar aggregates, as opposed to the wide distribution of compactness factors seen for neutral aggregates.

## ACKNOWLEDGMENTS

This work is supported by the National Science Foundation under Grant No. 0847127.